\documentclass[sigconf,natbib=true,anonymous=false]{acmart}

\usepackage{amssymb}
\usepackage{array}
\usepackage{url}
\usepackage{dblfloatfix}
\usepackage{booktabs}
\usepackage{amsmath}
\usepackage{tabularx}
\usepackage{colortbl}
\usepackage{mathtools}

\usepackage{graphics}
\usepackage{epsfig}
\usepackage{multirow}
\usepackage{flushend}

\usepackage{url}
\usepackage{tikz}
\usepackage{enumitem}
\usepackage{multicol}

\usepackage{verbatim}
\usepackage{float}
\usepackage{subcaption}

\usepackage{algorithm}
\usepackage{algpseudocode}

\usepackage{marginnote}
\usepackage{xspace}
\usepackage{balance}

\usepackage{todonotes}
\usepackage{xcolor}

\newcommand{\squishlist}{
 \begin{list}{$\bullet$}
  { \setlength{\itemsep}{0pt}
     \setlength{\parsep}{1pt}
     \setlength{\topsep}{1pt}
     \setlength{\partopsep}{0pt}
     \setlength{\leftmargin}{1.5em}
     \setlength{\labelwidth}{1em}
     \setlength{\labelsep}{0.5em} } }

\newcommand{\squishend}{
  \end{list}  }

\author{Iain Mackie}
\affiliation{
  \institution{University of Glasgow}
}
\email{i.mackie.1@research.gla.ac.uk}

\author{Ivan Sekulić	}
\affiliation{
  \institution{Università della Svizzera italiana}
}
\email{ivan.sekulic@usi.ch}

\author{Shubham Chatterjee}
\affiliation{
  \institution{University of Glasgow}
}
\email{shubham.chatterjee@glasgow.ac.uk	}

\author{Jeffrey Dalton}
\affiliation{
  \institution{University of Glasgow}
}
\email{jeff.dalton@glasgow.ac.uk}

\author{Fabio Crestani}
\affiliation{
  \institution{Università della Svizzera italiana}
}
\email{fabio.crestani@usi.ch}




\renewcommand\footnotetextcopyrightpermission[1]{} 

\begin{document}
\fancyhead{}

\title{GRM: Generative Relevance Modeling Using Relevance-Aware Sample Estimation for Document Retrieval}

\begin{abstract}

Recent studies show that Generative Relevance Feedback (GRF), using text generated by Large Language Models (LLMs), can enhance the effectiveness of query expansion. However, LLMs can generate irrelevant information that harms retrieval effectiveness. To address this, we propose Generative Relevance Modeling (GRM) that uses Relevance-Aware Sample Estimation (RASE) for more accurate weighting of expansion terms. Specifically, we identify similar real documents for each generated document and use a neural re-ranker to estimate their relevance. Experiments on three standard document ranking benchmarks show that GRM improves MAP by 6-9\% and R@1k by 2-4\%, surpassing previous methods.

\end{abstract}

\maketitle

\section{Introduction}
\label{sec:intro}



\begin{figure}[h!]
    \centering
    \setlength{\belowcaptionskip}{-7pt}
    \includegraphics[scale=0.16]{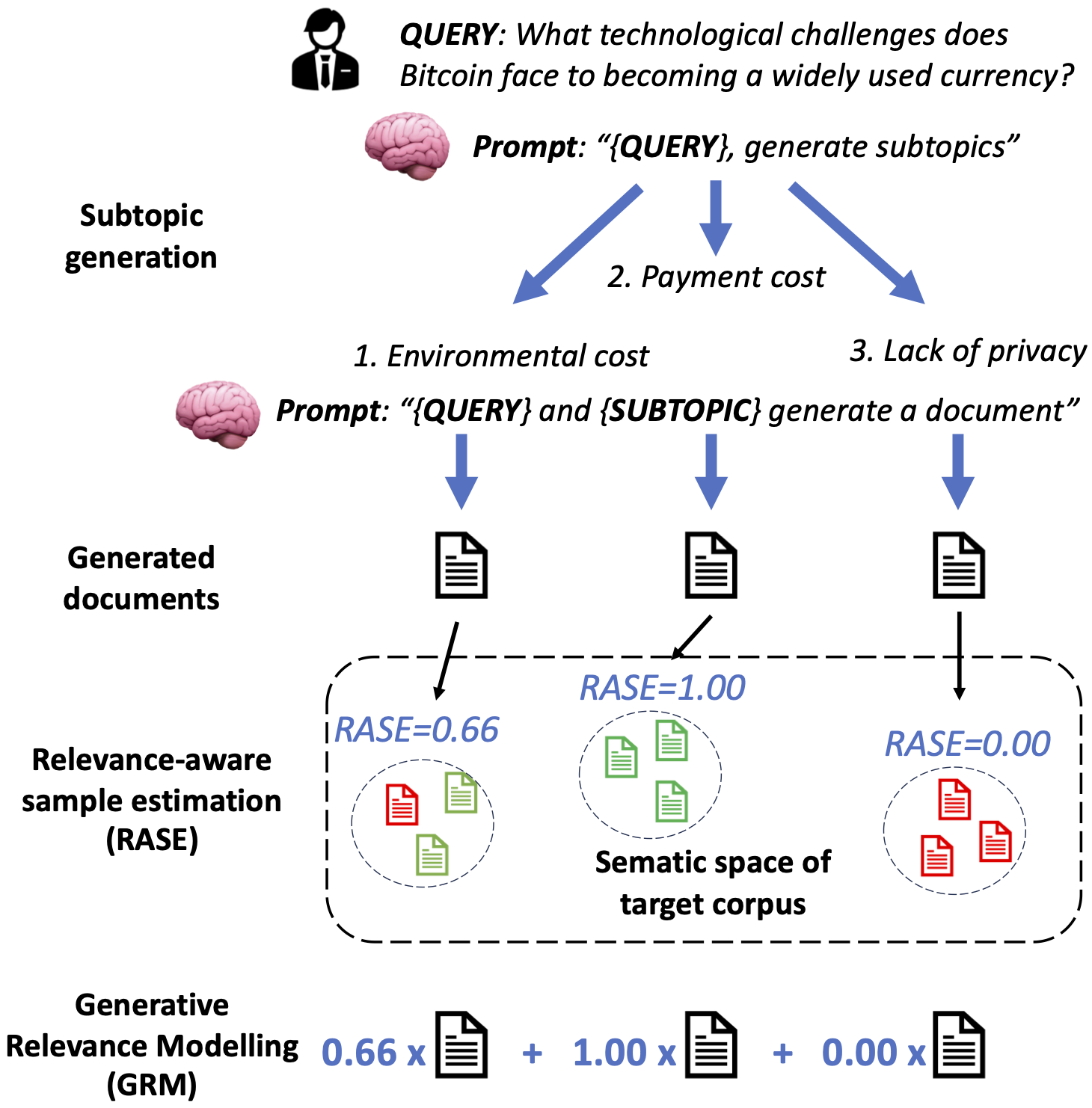}
    \caption{Generate Relevance Modeling (GRM) through Relevance-Aware Sample Estimation (RASE). This approach allows a diverse range of generated documents to be weighted within an expansion model based on the estimated relevance of semantically similar documents from the collection.}
    \label{img:GRM}
\end{figure}

The classical approach to addressing vocabulary mismatch~\cite{belkin1982ask} has been through Pseudo-Relevance Feedback (PRF), where the query is expanded with terms derived from the top-$k$ documents in a feedback set~\cite{belkin1982ask, abdul2004umass, zhai2001model, metzler2007latent, metzler2005markov}. Recent research on Generative-Relevance Feedback (GRF)~\cite{mackie2023generative} reveals that Large Language Models (LLMs) are capable of producing textual content that provides effective terms for query expansion. Nevertheless, LLMs are subject to generating ``hallucinations'' (text that isn't contained within the relevant documents), which is a considerable drawback. To address this issue, we propose the Generative Relevance Model (GRM) (see Figure ~\ref{img:GRM}) which leverages generated documents within a relevance modeling framework. However, we innovatively estimate the relevance of these generated documents based on their semantic similarity to relevant documents within the target collection, effectively addressing the shortcomings of the LLM-based generative expansion.


We focus on ``complex'' topics ~\cite{mackie2021deep, bolotova2022non} that demand context, reasoning ~\cite{Voorhees_TREC2004_robust}, and understanding of multiple concepts  ~\cite{mackie2022codec}, for example, ``challenges Bitcoin faces to become a widely accepted currency''. Our proposed GRM generates documents covering various subtopics related to the information need. This strategy essentially ``dissects'' the query into its constituent parts, which aids in better understanding the query itself. This, in turn, can improve the system's ability to retrieve relevant documents. However, there is a significant variance in the effectiveness of retrieval based on the selected documents for query expansion (see Section ~\ref{sec:results}). 
Initial experiments show that directly ranking the generated documents 
is ineffective due to the ranking model's unawareness of the target collection's documents. Therefore, we introduce Relevance-Aware Sample Estimation (RASE) that links our generated documents to similar ``real documents'' in the target collection. This method assists in weighing our GRM query expansion.


We experiment on TREC Robust04~\cite{Voorhees_TREC2004_robust} and CODEC~\cite{mackie2022codec} document ranking datasets.
We generate 50 documents per topic to give a reasonable sample and observe a significant variance in expansion effectiveness.
Specifically, if we could select the best document for query expansion, Recall@1k would be 0.83, while the worse generated document would result in a Recall@1k of 0.59.
Building upon this analysis, we show that GRM combined with RASE using a neural re-ranker~\cite{nogueira2020document} shows significant improvements over prior generative methods~\cite{mackie2023generative} and achieves new state-of-the-art results.

\textbf{Our contributions are as follows}: 

\begin{itemize}[leftmargin=*]
    \item We demonstrate LLMs can generate documents addressing subtopics of complex information needs. However, the effectiveness of expansion varies significantly depending on the document selected.
    \item We present GRM, a novel approach for modeling the relevance of generated documents based on the estimated relevance of semantically similar documents from the collection.   
    \item We show that our approach improves MAP by 6-9\% and Recall@1k by 2-4\% over prior state-of-the-art expansion methods.

\end{itemize}
\section{Related Work}
\label{sec:related-word}

Users typically express their information needs in the form of query, which is often under-specified or suffer from a lexical mismatch~\cite{belkin1982ask}. Although several methods have been developed to address this issue, query expansion remains a notable solution. 
Query expansion involves expanding the user's original query with terms more accurately representing the underlying information need~\cite{rocchio1971relevance}. These methods extract potentially useful terms from either explicit user feedback or, more commonly, through a pseudo-relevance feedback approach. Some well-known examples include vector space model feedback (like Rocchio~\cite{rocchio1971relevance}), query likelihood model (like RM3 expansion~\cite{abdul2004umass}), KL expansion~\cite{zhai2001model}, Relevance Modeling~\cite{metzler2005markov}, and Latent Concept Expansion~\cite{metzler2007latent}. Alternatively, query expansion could be achieved by extracting terms from a knowledge base~\cite{meij2010conceptual, xiong2015query, dalton2014entity}.

Recent advancements in Large Language Models (LLMs) have spurred progress across multiple IR research directions~\cite{yates2021pretrained}, including query rewriting~\cite{vakulenko2021question, wu-etal-2022-conqrr}, context and facet generation~\cite{macavaney2021intent5,hyde, liu2022query}, query-specific reasoning~\cite{pereira2023visconde, ferraretto2023exaranker}, and dataset generation~\cite{bonifacio2022inpars}. Additionally, researchers have explored dense vector representations for PRF~\cite{li2022improving}, leading to models like ANCE-PRF~\cite{yu2021improving}, ColBERT PRF~\cite{wang2022colbert}, and SPLADE-based PRF~\cite{lassance2023naver}. However, these efforts rely on documents retrieved in response to the original query, while our study highlights the effectiveness of multiple LLM-generated documents as context for query expansions.

Recent studies on Generative-Relevance Feedback (GRF) ~\cite{mackie2023generative} reveal that LLMs, specifically GPT-3 ~\cite{brown2020language}, can generate query-specific text independent of first-pass retrieval to enhance query expansion. This approach has been expanded \cite{mackie2023generativeArXiv} to dense and learned sparse retrieval paradigms, where generated content is encoded into vectors for query contextualization.
Our proposed approach varies in several ways. First, we propose a multi-turn prompting strategy to generate diverse documents based on subtopics. Second, we address the issue of ``hallucinations'' in generated content, where the generated text does not align with text within relevant documents from the collection. Specifically, we introduce our unique relevance-aware sample estimation technique to ensure that the most relevant generated documents are used for query expansion.

\section{Approach: GRM with RASE}
\label{sec:method}

In this section, we formally define our approach of generative relevance modeling (GRM) using relevance-aware sample estimation (RASE). We focus on the document retrieval task: Given an information need $Q$, retrieve a ranked list of relevant documents, 
$[\mathcal{D}_1, \mathcal{D}_2, \dots , \mathcal{D}_N]$ from a target collection. 

Specifically, we first use an LLM to generate a feedback set of synthetic documents, $[\mathcal{D}_{LLM}^1, \mathcal{D}_{LLM}^2, \dots, \mathcal{D}_{LLM}^K]$ for query expansion. To mitigate the variance in query expansion effectiveness based on the generated documents used within our relevance model, we propose Relevance-Aware Sample Estimation (RASE). This novel method estimates the relevance of generated documents based on the estimated relevance of semantically similar ``real'' documents from the target collection. The underlying assumption is that generated documents that resemble relevant documents from the collection should provide superior feedback signals. This method aids in identifying useful generated documents, thus reducing the impact of hallucinations and non-relevant terms within our query expansion. Together, these techniques aim to improve document retrieval by generating more diverse and relevant query expansions, mitigating the impact of less useful or off-topic synthetic content.


\paragraph{\textbf{Document Generation.}}



A complex query typically has multiple facets or subtopics. For example, different subtopics for the query ``What technological challenges does Bitcoin face to becoming a widely used currency?'' could be ``environmental cost'', ``payment costs'', or ``lack of privacy''. To provide a more comprehensive view of the information related to the query, we generate documents that cover different subtopics. Specifically, given the initial query $Q$, we utilize ChatGPT~\cite{brown2020language} with chain-of-thought reasoning~\cite{weichain} to first generate $K$ subtopics for the query $Q$. Then, we prompt the LLM to generate documents based on these subtopics, $\mathcal{D}_{LLM} = \{\mathcal{D}^1_{LLM},\mathcal{D}^2_{LLM}, \dots , \mathcal{D}^K_{LLM}\}$, where $\mathcal{D}^i_{LLM}$, the $i$-th LLM-generated document, covers the $i$-th subtopic for the query. This strategy helps avoid missing relevant documents that might be overlooked if we only focus on one aspect of the query. To provide a balance between depth (exploring a subtopic in more detail through multiple documents) and breadth (covering multiple subtopics), we perform this generation $G$ times, giving us $N$ ($K \times G$) diverse query-specific generated documents.



\paragraph{\textbf{Generative Relevance Model.}}
Our GRM builds upon prior work on query expansion, most notably RM3~\cite{abdul2004umass}. In this framework, we presume our generated documents, denoted as $\mathcal{D}_{LLM}$, are relevant and thereby define them as our relevant set $\mathcal{R}$. Our key task is to estimate the probability of observing a word $w$ given the relevance set, i.e., $P(w|\mathcal{R})$. Formally, we compute this as:

\begin{equation}
\centering
\label{eq:rm3-gen}
P(w|\mathcal{R}) = \sum_{D_{LLM} \in \mathcal{R}} P(w|D_{LLM}) \frac{P(Q|D_{LLM})}{\sum_{D'{LLM} \in \mathcal{R}} P(Q|D'{LLM})}
\end{equation}

In this equation, $P(Q|D_{LLM})$ is the query likelihood score, which quantifies the relevance of a document $D_{LLM}$ to the original query $Q$. This score is computed via the Relevance-Aware Sample Estimation (RASE) approach, discussed next.

\paragraph{\textbf{Relevance-Aware Sample Estimation.}}
RASE focuses on estimating the query likelihood score, $P(Q|D_{LLM})$, by modeling the relevance of documents in a collection that are similar to the generated document $D_{LLM}$. Specifically, for a given document collection $C$, we identify a subset of documents $\mathcal{D}_C = [\mathcal{D}_{C}^1, \mathcal{D}_{C}^2,\dots,\mathcal{D}_{C}^K]$ that are closest to $D_{LLM}$ according to a similarity function $\psi(D_{LLM}, C)$. We adopt the BM25 metric~\cite{robertson1994some} to measure this similarity, as inspired by recent work~\cite{macavaney:cikm2022-adaptive}. Next, we compute $P(Q|D_{LLM})$ using the relevance signals from $P(Q|D_C^i)$, operationalized via a Discounted Cumulative Gain (DCG) approach~\cite{jarvelin2002cumulated}:

\begin{equation}
\centering
\label{eq:query-like}
P(Q|D_{LLM}) = P(Q|D_{C}^i) + \sum_{i=2}^{K} \frac{P(Q|D_{C}^i)}{log_{2}(i)}
\end{equation}

Here, the DCG technique enables us to integrate a variety of relevance estimation models, from simpler methods like BM25 to more complex neural re-rankers like MonoT5~\cite{nogueira2020document}. Additionally, we can estimate an upper bound on our relevance estimation by leveraging query relevance judgments (``gold estimation''). In our experiments, we find that MonoT5 significantly improves the performance of our generative expansion methods~\cite{mackie2023generative}.

\section{Experimental Setup}
\label{sec:exp-setup}

\paragraph{\textbf{Datasets.}}



We conduct our experiments on two document ranking datasets. \textbf{CODEC}~\cite{mackie2022codec} is designed based on the complex information needs of social science researchers. It consists of 42 challenging essay-style topics produced by domain experts such as economists, historians, and politicians. The dataset contains a focused web corpus of 770k long documents, including sources such as BBC, Reuters, Brookings, Forbes, and eHistory.

\textbf{TREC Robust04}~\cite{Voorhees_TREC2004_robust} is created to improve retrieval effectiveness by targeting poorly performing topics. It comprises 249 topics with short keyword "titles" and longer natural language "descriptions" queries. The newswire collection contains 528k long documents (TREC Disks 4 and 5), which include sources such as the FT, Congressional Record, Federal Register, and the LA Times.

\paragraph{\textbf{Indexing.}} For indexing the corpora, we use Pyserini version 0.16.0 \cite{lin2021pyserini}, removing stopwords and using Porter stemming. Details of the hyperparameter tuning are provided with each method. We use cross-validation and optimize Recall@1000 on standard folds for Robust04~\cite{huston2014parameters} and CODEC~\cite{mackie2022codec}. 

\paragraph{\textbf{Evaluation.}} We evaluate our system runs up to a depth of 1,000, with the primary measure being Recall@1k. This choice is made due to the importance of recall-oriented evaluation in initial retrieval. We also include measures such as nDCG and MAP for additional comparison and to assess precision. All evaluations are conducted using ir-measures~\cite{macavaney2022streamlining}, and we use a paired t-test with a 95\% confidence interval to determine statistical significance.


\paragraph{\textbf{Document Generation.}}
We use the GPT-3 Chat API~\cite{brown2020language} for our document generation.
Specifically, we use the \texttt{gpt-3.5-turbo} model on Chat mode with parameters: $temperature=0.7$, $top\_p=1.0$, $frequency\_penalty=0.0$, $presence\_penalty=0.0$, and maximum length of 512.
We 1-shot prompt ChatGPT to create five subtopics ($K=5$) before we generate documents based on these subtopics.
We repeat this process $G=10$ times to create a reasonable set of 50 diverse generated documents per topic. We will release all prompts and generated documents for reproducibility.

\paragraph{\textbf{Retrieval and Expansion.}}
For the preliminary run, we employ a fine-tuned BM25 model~\cite{robertson1994some}. Specifically, we tune the $k_1$ parameter within the range of 0.1 to 5.0, with a step size of 0.2, and the $b$ parameter within the range of 0.1 to 1.0, with a step size of 0.1.

For RASE , we use prior work on document-to-document similarities~\cite{macavaney:cikm2022-adaptive} as a reference. We apply the tuned BM25 model for the document similarity function $\psi$, treating the generated document as a query. Additionally, we tune the number of documents retrieved from the target collection, ranging from 10 to 100 with a step of 10.


\paragraph{\textbf{Relevance Estimate Functions.}}
For our relevance estimate function, $P(Q|D_C)$, we use the following four formulations and normalize scores:

\begin{itemize}[leftmargin=*]

\item \textbf{GRM-Uniform}: We set the relevance estimation to 1.0 to show the impact of other methods. 

\item \textbf{GRM-BM25}: Use the tuned BM25~\cite{robertson1994some} model.

\item \textbf{GRM-T5}: We use the T5-3B~\cite{nogueira2020document} re-ranker and max-passage aggregate to calculate the document scores.

\item \textbf{GRM-Gold}: We use scaled relevance judgments to show oracle.

\end{itemize}

\paragraph{\textbf{Other GRM Hyperparameters}}. We tune the remaining GRM hyperparameters: the number of feedback docs ($fb\_docs$: 5 to 95 with a step of 10), the number of feedback terms ($fb\_terms$: 5 to 50 with a step of 10), the interpolation between the original terms and generative expansion terms ($original\_query\_weight$: 0.1 to 0.9 with a step of 0.1). The tuning methodology is the same as BM25, BM25 with RM3 expansion and GRF to make them directly comparable.

\paragraph{\textbf{Baselines.}} We compare our approach to the following systems:


\begin{enumerate}[leftmargin=*]

\item \textbf{BM25+RM3}~\cite{abdul2004umass}: 
For BM25, we tune $k1$ (0.1 to 5.0 with a step of 0.2) and $b$ (0.1 to 1.0 with a step of 0.1).
Then for RM3 expansion: $fb\_terms$ (5 to 95 with a step of 5), $fb\_docs$ (5 to 50 with a step of 5), and $original\_query\_weight$ (0.1 to 0.9 with a step of 0.1). 

\item \textbf{CEQE}~\cite{naseri2021ceqe}: Utilizes query-focused vectors for query expansion. We use the CEQE-MaxPool runs provided by the author. 

\item \textbf{SPLADE+RM3}: We use SPLADE~\cite{formal2021splade} with RM3~\cite{abdul2004umass} expansion. We use Pyserini's~\cite{lin2021pyserini} ``impact'' searcher, max-passage aggregatation, and \texttt{naver/splade-cocondenser-ensembledistil}. We tune $fb\_docs$ (5,10,15,20,25,30), $fb\_terms$ (20,40,60,80,100), and $original\_query\_weight$ (0.1 to 0.9 with a step of 0.1).

\item \textbf{TCT+PRF}~\cite{Li2021PseudoRF}: Roccio PRF using ColBERT-TCT~\cite{lin2021batch}. We employ max-passage approach with TCT-ColBERT-v2-HNP checkpoint. We tune Roccio PRF parameters: $depth$ (2,3,5,7,10,17), $\alpha$ (0.1 to 0.9 with a step of 0.1), and $\beta$ (0.1 to 0.9 with a step of 0.1).  

\item  \textbf{GRF}~\cite{mackie2023generative} : Expands the query based on the language model from text aggregated across multiple LLM-generation subtasks. We use the full GRF method and the \textbf{GRF-News} variant for comparison, using the runs provided by the author. 

\end{enumerate}

\section{Results and Analysis}
\label{sec:results}







\begin{table*}[t]
\caption{Document retrieval effectiveness of GRM with different RASE methods (Uniform, BM25, T5, Gold). ``+'' indicates significant improvements against the full GRF and \textit{bold} depicts the best system. }
\label{tab:main}
\centering
\begin{tabular}{ll|lll|lll|lll|}
\cline{3-11}
                                                    &                    & \multicolumn{3}{c|}{Robust04 - Titles}                   & \multicolumn{3}{c|}{Robust04 - Descriptions}                 & \multicolumn{3}{c|}{CODEC}                                   \\ \cline{3-11} 
                                                    &                    & MAP                & nDCG               & R@1k           & MAP                & nDCG               & R@1k               & MAP                & nDCG               & R@1k               \\ \hline
\multicolumn{1}{|l|}{\multirow{4}{*}{\textit{PRF}}} & TCT+PRF            & 0.274              & 0.541              & 0.684          & 0.245              & 0.493              & 0.628              & 0.239              & 0.532              & 0.757              \\
\multicolumn{1}{|l|}{}                              & SPLADE+RM3         & 0.248              & 0.518              & 0.703          & 0.268              & 0.535              & 0.715              & 0.216              & 0.506              & 0.770              \\
\multicolumn{1}{|l|}{}                              & CEQE \cite{naseri2021ceqe}              & 0.310              & 0.579              & 0.764          & -                  & -                  & -                  & -                  & -                  & -                  \\
\multicolumn{1}{|l|}{}                              & BM25+RM3           & 0.292              & 0.571              & 0.777          & 0.278              & 0.551              & 0.750              & 0.239              & 0.530              & 0.816              \\ \hline
\multicolumn{1}{|l|}{\multirow{2}{*}{\textit{GRF}}} & GRF-News \cite{mackie2023generative}             & 0.287              & 0.571              & 0.745          & 0.274              & 0.557              & 0.717              & 0.270              & 0.573              & 0.828              \\
\multicolumn{1}{|l|}{}                              & GRF \cite{mackie2023generative}               & 0.307              & 0.603              & 0.788          & 0.318              & 0.605              & 0.776              & 0.285              & 0.585              & 0.830              \\ \hline
\multicolumn{1}{|l|}{\multirow{4}{*}{\textit{GRM}}} & GRM-Uniform (Ours) & 0.306              & 0.594              & 0.778          & 0.313              & 0.605              & 0.779              & 0.306$^+$          & \textbf{0.611$^+$} & \textbf{0.850$^+$} \\
\multicolumn{1}{|l|}{}                              & GRM-BM25 (Ours)    & 0.312              & 0.599              & 0.781          & 0.315              & 0.607              & 0.779              & 0.303$^+$          & 0.608$^+$          & 0.843              \\
\multicolumn{1}{|l|}{}                              & GRM-T5 (Ours)     & \textbf{0.327$^+$} & \textbf{0.615$^+$} & \textbf{0.796} & \textbf{0.342$^+$} & \textbf{0.631$^+$} & \textbf{0.805$^+$} & \textbf{0.309$^+$} & \textbf{0.611$^+$} & 0.848$^+$          \\ \cline{2-11} 
\multicolumn{1}{|l|}{}                              & GRM-Gold (Ours)    & 0.388$^+$          & 0.672$^+$          & 0.819$^+$      & 0.387$^+$          & 0.675$^+$          & 0.823$^+$          & 0.336$^+$          & 0.642$^+$          & 0.855$^+$          \\ \hline
\end{tabular}
\end{table*}

\begin{figure}
    \centering
    \setlength{\belowcaptionskip}{-7pt}
    \includegraphics[width=\columnwidth]{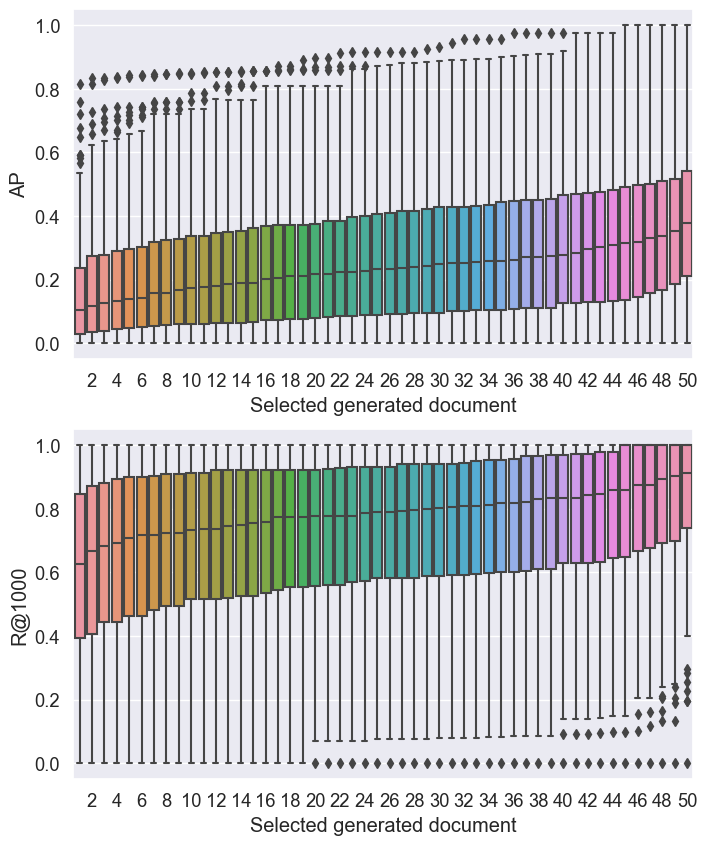}
    \caption{MAP and R@1000 boxplot of varying generated documents ordered by effectiveness on Robust04 titles, i.e. the worst generated document for expansion for each topic to the left (1) and best generated document to the right (50).}
    \label{fig:perf_branches}
\end{figure}

\paragraph{\textbf{RQ1: Does selection of generated documents impact query expansion effectiveness?}}

Figure~\ref{fig:perf_branches} displays how the choice of generated documents used for expansion impacts retrieval effectiveness on Robust04 title queries. The boxplot represents the query effectiveness distribution if we select documents for expansion from the worst to the best (from left to right).

\textbf{Effectiveness based on selection quality.} The selection of generated documents significantly affects the overall effectiveness. For instance, MAP ranges from 0.21 for the worst-generated documents per query to 0.34 for the oracle-generated document. Similarly, the worst possible document results in an R@1000 of 0.59, whereas the best-generated document increases recall by 0.24 to 0.83.

\textbf{Query variance.} The boxplots show that the effectiveness differs greatly based on the query, suggesting high variance even with constant selection quality. For example, some queries achieve MAP of 0.0 (even with the Oracle document), indicating no relevant documents were retrieved. Conversely, some queries achieve MAP of 0.81 and R@1000 of 1.0 even with the worst possible document. This demonstrates that generating multiple documents doesn't necessarily alleviate the difficulty of hard queries.

These findings strongly suggest the potential of LLMs for generating documents for query expansion, especially for complex topics. However, they also highlight an issue: the effectiveness of different generated documents used for query expansion can vary dramatically. While some documents align with the content in relevant documents from the target collection, others do not. In the following section, we propose a solution to weigh generated documents more effectively based on their semantic similarity to the collection's relevant documents.

\paragraph{\textbf{RQ2: Does generative relevance modeling improve retrieval effectiveness?}}

Table~\ref{tab:main} presents document retrieval effectiveness on Robust04 and CODEC datasets using different relevance estimate functions (outlined in Section \ref{sec:exp-setup}). 

Interestingly, GRM-Uniform performs similarly to GRF on Robust04, suggesting that blindly generating diverse subtopics may not always improve query expansion for difficult, specific topics. However, using BM25 for relevance estimation leads to small but consistent gains over Uniform, although still not significant gains.

Utilizing a neural re-ranker~\cite{nogueira2020document}, such as in GRM-T5, reveals that better RASE can more effectively weigh beneficial generated documents for query expansion. Specifically, this results in significant gains across all measures on Robust04 descriptions and on MAP and nDCG on Robust04 titles over GRF. In fact, GRM-T5 outperforms all PRF models. Our results also indicate that an Oracle relevance estimation could further boost performance by for MAP 13-19\% and by 2-3\% for R@1k.

On CODEC, all GRM methods significantly outperform GRF. Since CODEC's complex topics often encompass multiple subtopics, our diverse subtopic-driven prompting approach greatly enhances effectiveness for GRM-Uniform. Relevance estimation via BM25 offers little improvement in R@1k, but T5 significantly improves all measures. Interestingly, while GRM-Gold boosts precision in MAP and nDCG, it shows less improvement in recall.

Overall, our results demonstrate that using a neural re-ranking for RASE in GRM significantly improves effectiveness over GRF methods, leading to an increase of 6-9\% in MAP, 2-4\% in nDCG, and 2-4\% in R@1k. This highlights the promising potential of GRM for state-of-the-art document retrieval across multiple datasets.

\section{Conclusions}
\label{sec:conclusion}

In conclusion, this work presents Generative Relevance Modeling (GRM), a novel approach to document retrieval that leverages large language models to generate diverse synthetic documents covering a wide array of subtopics related to an initial query. Our study on challenging datasets demonstrates that our approach significantly enhances document retrieval effectiveness.
We found a considerable variation in retrieval effectiveness, with some generated documents providing near-perfect context while others veered the query off-topic. To address this, we introduced Relevance-Aware Sample Estimation (RASE) within GRM to estimate the relevance of generated documents based on their similarity to relevant documents from the target collection.
We show that GRM with RASE significantly improves upon traditional generative expansion methods, increasing MAP by 6-9\% and R@1k by 2-4\%.
In summary, our work underscores the potential of using LLMs in information retrieval systems and encourages further research on improving the generation and weighting process of synthetic documents.

\section{Acknowledgements}
\label{sec:ack}

This work is supported by the 2019 Bloomberg Data Science Research Grant and the Engineering and Physical Sciences Research Council grant EP/V025708/1. 




\bibliographystyle{ACM-Reference-Format}
\balance
\bibliography{foo}

\end{document}